\documentclass[aps,prl,letterpaper,11pt,twoside,tightenlines,nofootinbib,showpacs,preprint,onecolumn]{revtex4}
\usepackage{graphicx}
\usepackage[sort&compress]{natbib}
\usepackage{latexsym}
\usepackage{epsfig}
\begin{document}
\newcommand{\eg}{{\it e.g.}}
\newcommand{\etal}{{\it et. al.}}
\newcommand{\ie}{{\it i.e.}}
\newcommand{\be}{\begin{equation}}
\newcommand{\dd}{\displaystyle}
\newcommand{\ee}{\end{equation}}
\newcommand{\bea}{\begin{eqnarray}}
\newcommand{\eea}{\end{eqnarray}}
\newcommand{\bef}{\begin{figure}}
\newcommand{\eef}{\end{figure}}
\newcommand{\bce}{\begin{center}}
\newcommand{\ece}{\end{center}}
\def\lsim{\mathrel{\rlap{\lower4pt\hbox{\hskip1pt$\sim$}}
    \raise1pt\hbox{$<$}}}         
\def\gsim{\mathrel{\rlap{\lower4pt\hbox{\hskip1pt$\sim$}}
    \raise1pt\hbox{$>$}}}         

\centerline{Unified approach to growth and aging in biological, technical and biotechnical  systems}
\vskip 20pt
\centerline{ Ph. Blanchard$^{(a)}$ and P.Castorina$^{(c,b)}$}
\vskip 20pt
\centerline{$^{(a)}$ Fakult$\ddot a$t f$\ddot u$r Physik, Universit$\ddot a$t Bielefeld, Germany}
\vskip 10pt
\centerline{$^{(b)}$ Dipartimento di Fisica, Universita' di Catania, Italy}
\vskip 10pt
\centerline {$^{(c)}$ INFN-Catania, Italy}

\vfill
\eject
\section{Abstract}
\vskip 30pt
Complex systems, in many different scientific sectors, show coarse-grain properties with simple growth laws with respect to fundamental microscopic algorithms. We propose a classification scheme of growth laws which includes human aging, tumor (and/or tissue) growth, logistic and generalized logistic growth and the aging of technical devices.  The proposed classification permits to evaluate the aging/failure of combined new bio-technical "manufactured products", where part of the system evolves in time according to biological-mortality laws and part according to technical device behaviors. Moreover it suggests a direct relation between the mortality leveling-off for humans and technical devices and the observed small cure probability for large tumors. 
\vskip 30pt
Pacs {89.75.-k}
\vskip 30pt
Keywords : growth laws, aging, tumor growth
\vskip 30pt
Corresponding author : Paolo Castorina, paolo.castorina@ct.infn.it

\vfill
\eject

\section{Background}

Complex systems with millions of interacting elementary parts are often considered
computationally irreducible \cite{wolf1,wolf2} which means that the only way to decide about their evolution
is to let them evolve in time.

On the other hand, there is an impressive number of experimental verifications, in many different scientific sectors, 
that  coarse-grain properties of systems, with simple laws with respect to fundamental microscopic alghoritms,
emerge at different levels of magnification providing important tools for explaining and predicting new phenomena.

In this respect, a priori unrelated systems show similar emergent properties and if an unexpected effect is found experimentally in a field, a similar effect, "mutatis mutandis", should also be sought in similar  experiments in other fields. Therefore a useful tool to greatly facilitate the cross fertilization among different fields of research is a general classification  of growth laws \cite{castorina}.

A very important example is the Gompertz law  (GL) \cite{gompertz} which applies to human mortality tables ( i.e. aging) and  tumor growth \cite{steel,wh,norton}. 

In general, a growth problem is characterized by a function $f(t)$, which describes the time evolution of some macroscopic quantity,
and by the specific  rate , $\alpha$,  defined as $(1/f)(df/dt)=\alpha(t)$. In the GL $\alpha$ has an exponential dependence on time:
\be
(1/f)(df/dt)=\alpha(t)= a e^{bt},
\ee
where $a$ and $b$ are constants. In aging $f(t)$  indicates the survival probability; while with regards to tumor growth it corresponds to the number of cells $N(t)$ ( depending on the specific case $a$ and $b$ can be positive or negative).

For technical devices the specific rate of the survival probability has a  power-law  time behavior 
\be
(1/f)(df/dt) = \alpha(t) = a t^n,
\ee 
with $n>1$, called  Weibull law (WL) \cite{barlow,rigdon}. The analogy with the biological systems is intriguing (for clarity, as necessary, one defines the specific rate $\alpha_h(t)$ for the human mortality,  $\alpha_f(t)$ for the technical systems and $\alpha_c(t)$ for tumor growth) and deeper than the similarity between eq.(1) and eq.(2).  

Indeed, many independent analyses of experimental data on humans and animals suggest that at advanced ages (more than 85-90 years for humans) there is
a deceleration in mortality \cite{gavrilov1,vaupel,olshansky} :  in the large range 20 - 85 years for humans the mortality rate is well described by the Gompertz law and then there is a late-life mortality (although a definite conclusion has yet to be reached \cite{gavri0}). A similar trend is observed for technical devices \cite{economos}, confirming the analogy between biological and technical systems.

The understanding of aging and of late-life mortality is still an open problem and many interesting models have been proposed to explain the similar  behavior in metabolic systems and in technical devices \cite{gavri}. Moreover, a unifying language for the description of performance  of metabolic and technical production and distribution has been recently suggested \cite{hutt} to implement the idea that the robustness of metabolic systems with respect to enviromental changes could represent a useful model for technical systems. 

In this letter, rather than focusing on specific models, we shall address the generalization of the classification scheme of growth laws  to  include human aging, tumor (and/or tissue) growth, logistic and generalized logistic growth and the aging of technical devices. We shall consider two applications of the proposed approach: a) a method to evaluate the aging/failure of combined new bio-technical "manufactured product", where part of the system evolves in time according to biological-mortality laws and part is a technical device; b) an interpretation of the  "tumor size effect", i.e. the small cure probability for large tumor\cite{stanley,bentzen,huchets}, in analogy with the late-life mortality in aging.

\section{Results}

Let us start with the general classification scheme.  It turns out that a classification of the growth laws according to the simple equation $(1/f)(df/dt)=\alpha(t)$ is obtained by considering the power expansion in $\alpha$ of the function ( see ref. \cite{castorina} for details)
\be
\Phi(\alpha) = \frac{d\alpha}{dt} = \Sigma_i b_i \alpha^i \phantom{..} i=0,1,2...
\ee
which for $b_0=0$ and $b_i=0$ for $i>1$ gives a time independent specific rate $\alpha_0$ and therefore an exponential growth; for
$b_0 \neq 0$ and $b_i=0$ for $i>1$ describes a linear time dependent specific rate and again an exponential growth; at the first order in $\alpha$, for
$b_0= 0$, $b_1 \neq 0$  and $b_i=0$ for $i>1$, reproduces an exponential time behavior of the specific growth and therefore the GL;
the second order term , $O(\alpha^2)$, for $b_0= 0$, $b_1,b_2 \neq 0$  and $b_i=0$ for $i>2$  generates the logistic and generalized logistic growth.

The feedback effect,  that is the dependence of the specific growth rate $\alpha$   on the function $f(t)$,
 can be easily derived by  the temporal behaviour of the specific rate. For the GL for a growing number of cells, $N(t)$, one has the well known logarithmic non linearity,
\be
\frac{1}{N(t)} \frac{dN(t)}{dt} = a - b \ln{\frac{N(t)}{N_0}} = b \ln{\frac{N_\infty}{N(t)}} \phantom{.....} Gompertz,
\ee
and  for the (generalized) logistic law one gets the typical power-law behavior 
\be
\frac{1}{N(t)} \frac{dN(t)}{dt} = c [1- (\frac{N(t)}{N_\infty})^\gamma]  \phantom{...}  gen. \phantom{.} logistic,
\ee
where $a,b,c,\gamma $ are constants and the carrying capacity,  $N_\infty$ ,corresponds to $\alpha=0$.

In order to describe technical devices, the previous classification scheme has to be generalized since the specific growth rate of Weibull law
has a power law dependence on time  which is not reproduced by eq.(3). The behavior $\alpha_f(t) \simeq t^n$,with $n$ positive integer,  corresponds to
terms O( $\alpha^{(n-1)/n})$ in the expansion of $\Phi(\alpha)$ and  therefore for a general classification  scheme of the specific growth/aging/failure  rate of  biological and technical systems one has to consider:
\be
\Phi(\alpha)= \Sigma_{n>2}^\infty c_n \alpha^{(n-1)/n} +  \Sigma_{n \ge 1} b_n \alpha^n
\ee 
Note that: a)  $0<(n-1)/n<1$ and the nth term in the power series in $\alpha^{(n-1)/n}$ tends for large $n$ to  $\alpha$, i.e. to the Gompertz law;
b) the term $b_0 \neq 0$, i.e. the exponential growth, has been neglected because one considers  the GL, the generalized logistic or more complex growth laws for the biological systems (there is no problem to include this term in the expansion) ; 3) the first sum in the expansion has fractional powers that recall a Puiseux expansion.

As a by-product of the proposed classification scheme one can easily evaluate the aging/failure of combined new bio-technical "manufactured products" by taking explicitely into account the mutual "interference" between the aging behavior of the biological part  and the failure of the technical one.
 The  "interference" effect strongly depends on the typical time scales  in the coefficients $c_n$ and $b_n$ in the previous expansion:
if the life-time of the technical device is much larger than the life-time of the biological part ( or viceversa) there is essentially no effect  \cite{pace}.

\begin{figure}
{{\epsfig{file=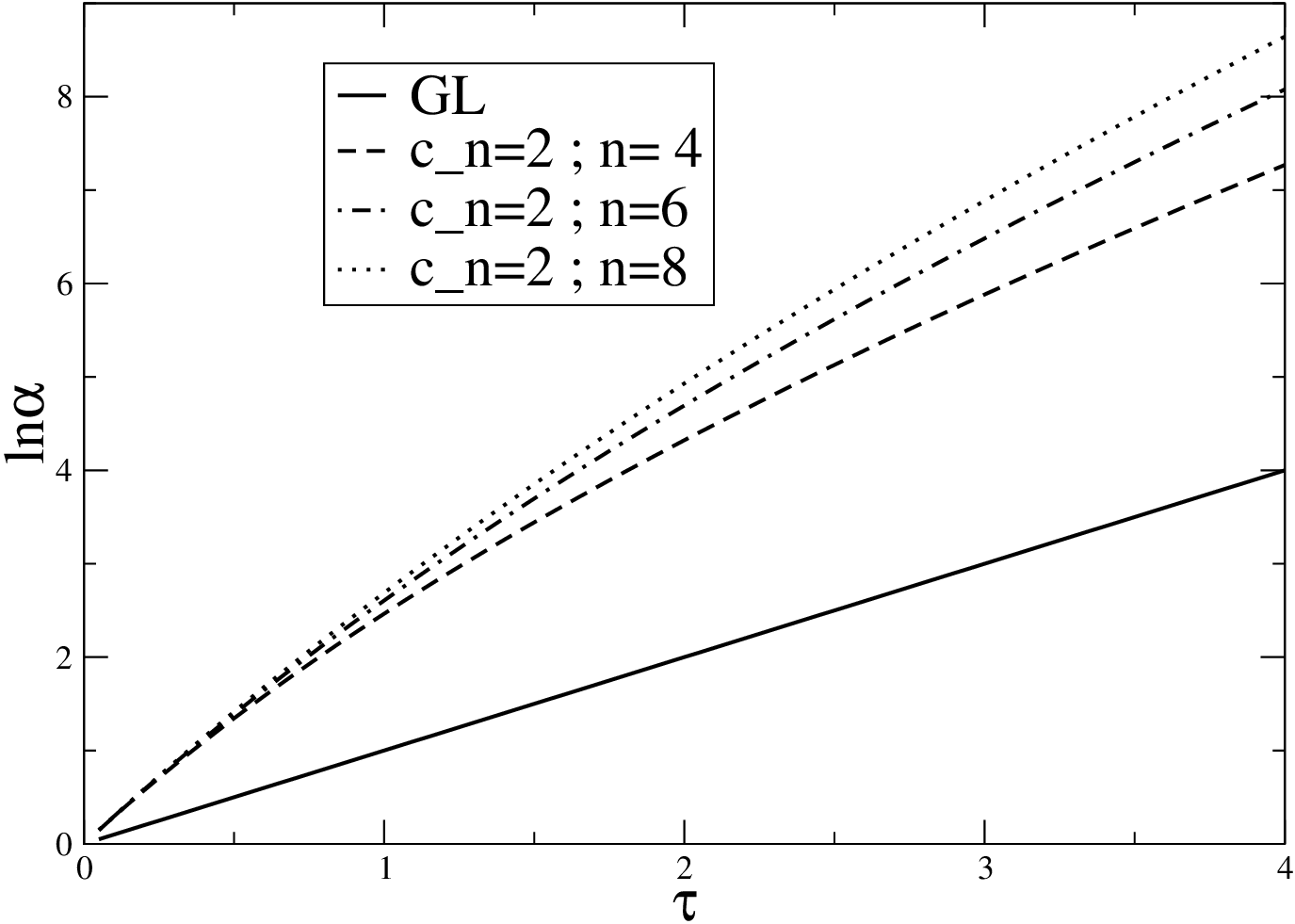,height=7.0 true cm,width=7.0 true cm, angle=0}}
\caption{
}}
\end{figure}

Let us first  consider   aging/failure of a combined  bio-technological "manufactured product", where part of the system evolves in time according to GL, i.e. the term $O(\alpha)$,  and the behavior of technical part  is described by  a single term $O(\alpha^{n-1/n})$,i.e. 
\be
\Phi(\alpha)=  c_n \alpha^{(n-1)/n} +b_1 \alpha
\ee
By introducing  dimensionless variables in time unit $1/b_1$, i.e. $\tau=b_1t$, $\bar \alpha = \alpha/b_1$ and $\bar c_n = c_n b_1^{-1-1/n}$, after simple calculations the time dependence of the specific rate is given by:
\be
e^{\tau} = \frac{\bar \alpha}{\bar \alpha_0} \frac{[1 + (\bar c_n)  \bar \alpha^{-1/n}]^n}{[1 + (\bar c_n) \bar \alpha_0^{-1/n}]^n}
\ee
where $\bar \alpha_0=\bar \alpha(\tau=0)$. Of course in the limit $c_n \rightarrow 0$ one recovers the GL and for $b_1 \rightarrow 0$ the Weibull one. By previous equation, for $\bar \alpha_0 = 1$, one obtains:
\be
\ln{\bar \alpha} = n \ln{[(1+\bar c_n) e^{\tau/n} - \bar c_n]}
\ee
which describes the combined effect of the two growth laws. The quantitative effect is depicted in figs. (1,2) where  the previous function is plotted for different values of $n$ at fixed $\bar c_n$ and for various values of  $\bar c_n$ at fixed $n$. 

\vskip5pt

\begin{figure}
{{\epsfig{file=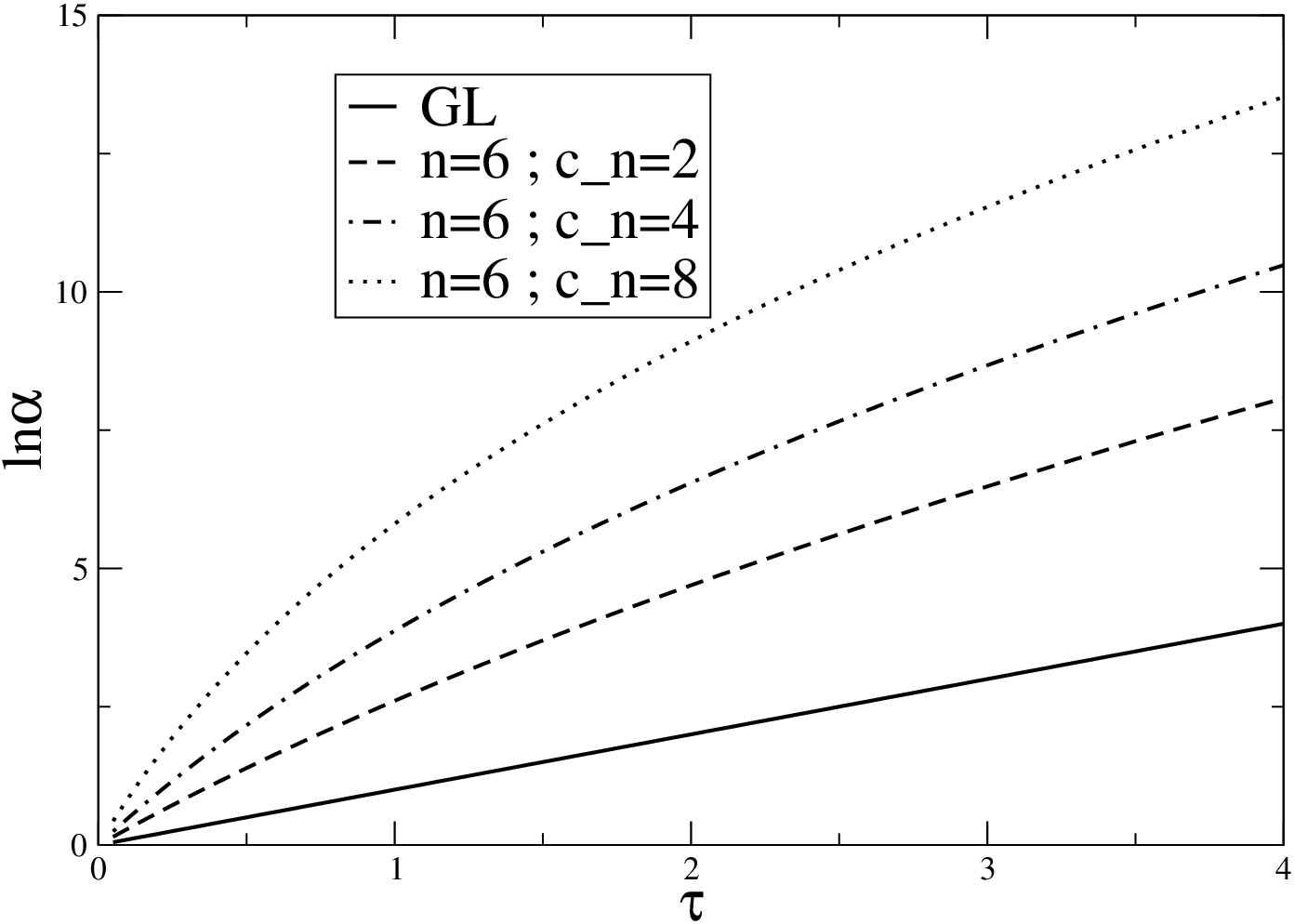,height=7.0 true cm,width=7.0 true cm, angle=0}}
\caption{
}}
\end{figure}

\vskip5pt 

The next step is to include the term $ b_2 \alpha^2$ in the expansion of $\Phi(\alpha)$ ($b_2$ is dimensionless) which corresponds to a generalized logistic evolution. As we shall see this term is crucial in understanding the late-life mortality effect.

 By repeating analogous calculations it turns out that
\be
\tau= ln(\bar \alpha /\bar \alpha_0) - \int_{\bar \alpha_0}^{\bar \alpha} dx \frac{b_2 + \bar c_n x^{-(1-n)/n}}{1+b_2 x + \bar c_n x^{-1/n}}  
\ee
In fig.3 is shown that the term $ b_2 \alpha^2$ completely changes the time evolution with respect to GL and/or WL producing a leveling-off of the specific rate.

Therefore the general expansion of $\Phi(\alpha)$ in eq.(6) can describe the aging/failure of any biological and technical system including the
leveling-off at late mortality which  is obtained by taking into account the term $O(\alpha^2)$ in $\Phi(\alpha)$, i.e. by the transition from the GL or WL  to a logistic type law \cite{hori}.

The proposed unification scheme suggests a  practical method to understand growth
patterns. Given a set of data on some growth process, the first step of the analysis is a fit in power of $\alpha$  of the derivative of the specific growth rate, i.e. of the function $\Phi(\alpha)$.
Therefore : a) if the best fit is linear, the growth is  a Gompertzian one; b) if the best fit is quadratic, look at the sign of the coefficients of the expansion.  For  $b_1 >0$ and $b_2<0 $ the growth is logistic (or generalized logistic) corresponding to a competitive dynamics; 
c) if the best fit indicates a fractional power the growth follows the WL. Of course, it is always possible to obtain a better agreement with data  by increasing the number of coefficients. However, should increasing the number of parameters indicate only a marginal improvement in the  description of  data one  concludes that the added terms in the expansion are irrelevant.

\section{Discussion and Conclusions}

Let us now consider the cross-fertilization among different sectors. 

As previously discussed, there is a deceleration of mortality in aging at late time which is described as a "transition" from a Gompertz law to a generalized logistic behavior. On the other hand, tumors evolve in time according to the GL. The obvious indications is to verify if a phenomenon corresponding to the deceleration of mortality,  i.e. a transition from the GL to a  power law, exists for  cancer growth at a later time.
As we shall see, this aspect has strong consequences on the therapy.

For tumor growth the $b_1 \alpha$ term gives the GL  in eq.(4) and the introduction of the $O(\alpha^2)$ term corresponds to the power law non-linear feedback in eq.(5). Therefore one has to investigate if at late-life of a tumor growth  there is such a modification in the dependence 
of the specific  growth rate on the cell number $N(t)$. Since direct informations " in vivo" are almost impossible, the question has to be addressed in  an indirect way by considering radiotherapy.

The radiotherapic tumor treatment consists in series of radiation doses at fixed time intervals. However tumors start to re-grow in the interval between two treatments : the re-growth during radiotherapy  is therefore an important clinical parameter \cite{kim} and  the probability of treatment benefit critically depends on the tumor re-growth pattern. 

The so called " tumor size effect" is a reduction of radiotherapeutic results for large tumors ( which , presumably, has grown since long time).
The dependence of the surviving fraction on the tumor volume was already observed by Stanley et al. in 1977 
in lung tumors \cite{stanley} and re-emphasized by Bentzen et al. and Huchet et al. in \cite{bentzen,huchets}.

The effect of re-growth rate on radiotherapy has been quantitatively investigated in ref. \cite{cast2} and the results clearly indicate that to understand the tumor size effect  the re-growth rate for large tumor has to follow a power law \cite{guiot} rather than the GL.

From this point of view the " tumor size effect" is a phenomenon which indicates that in late -time tumor growth there is a change from a GL specific rate to a power law behavior, corresponding  to the  deceleration in mortality at advanced age.  

One should conclude that such a common feature in aging and in failure in biological and/or technical systems should be considered as a
"bifurcation" or a "phase transition"  in the specific  growth rate at large time from GL or WL to a logistic or generalized logistic behavior.

In closing, the general expansion of $\Phi(\alpha)$ in eq.(6) can describe the growth/aging/failure of biological and technical systems and the transition to a different ("phase") specific growth rate at late-life could be a common feature of those systems independently on the microscopic dynamics.

\begin{figure}
{{\epsfig{file=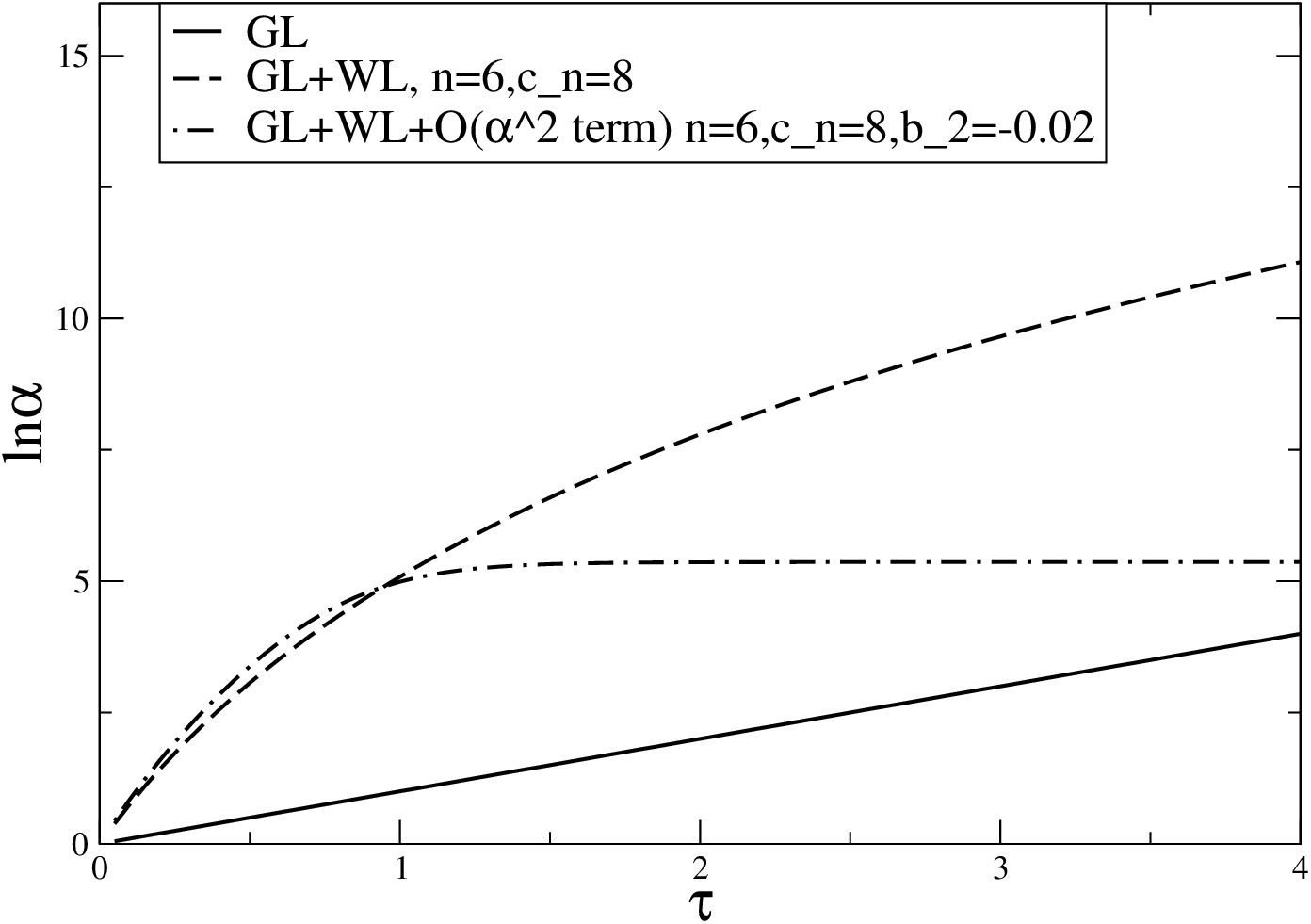,height=7.0 true cm,width=7.0 true cm, angle=0}}
\caption{
}}
\end{figure}

\vfill
\eject

\vfill
\eject

Captions
\vskip 30pt

Fig. 1 : Comparison of the GL, the WL and the combined effect for a biotechnical device for $\ln{\bar \alpha} $. $\tau=b_1t$ and the curves are for a fixed value of the coefficient $\bar c_n =2$ and different values of $n=4,6,8$.

\vskip 30pt

Fig. 2 : Comparison of the GL, the WL and the combined effect for a biotechnical device for $\ln{\bar \alpha} $. $\tau=b_1t$ and the curves are for $n=6$ and the coefficient
$\bar c_n =2,4,8$ 

\vskip 30pt

Fig. 3 : Comparison for $\ln{\bar \alpha} $ of the GL, the WL and the effects of $O(\alpha^2)$ term  for $n=6$,$\bar c_n=8$ and $b_2=-0.02$.

\vskip 30pt

\end{document}